\documentclass[12pt]{article}
\pdfoutput=1
\usepackage{amsmath, amssymb, amsfonts, amsthm, mathtools}
\usepackage{times} 
\usepackage{graphicx}
\usepackage{color, dsfont}
\usepackage{graphicx}
\usepackage{mathabx}
\usepackage{multirow}
\usepackage{setspace}
\usepackage{tabularx}
\usepackage[labelsep=period]{caption}
\usepackage{subcaption}
\usepackage[hidelinks,bookmarks=false]{hyperref}
\usepackage[page,title,titletoc,header]{appendix}
\usepackage[top=1in, bottom=1in, left=1in, right=1in]{geometry}
\usepackage{url,indentfirst,comment,titlesec,secdot,enumerate,psfrag,epsf}
\usepackage[style=authoryear-comp,sortcites=false,firstinits=true, backend=bibtex,natbib=true,hyperref=true,maxbibnames=99,maxcitenames=2,language = american]{biblatex}
\usepackage{lipsum}
\usepackage{sectsty}

\titleformat*{\section}{\large\bfseries\centering}
\titleformat*{\subsection}{\normalsize\bfseries}	

\newcolumntype{L}[1]{>{\raggedright\arraybackslash}p{#1}}
\newcolumntype{C}[1]{>{\centering\arraybackslash}p{#1}}
\newcolumntype{R}[1]{>{\raggedleft\arraybackslash}p{#1}}
\renewbibmacro{in:}{}
\sectionfont{\bfseries\large\raggedright}
\subsectionfont{\normalsize\raggedright}
\allowdisplaybreaks

\author{Olanrewaju Akande, Andr\'{e}s Barrientos, and Jerome P. Reiter \footnote{Olanrewaju Michael Akande is PhD Candidate, Department of Statistical Science, Duke University, Durham, NC 27708 (E-mail: \href{mailto:olanrewaju.@duke.edu}{olanrewaju.akande@duke.edu}); Andr\'{e}s Felipe Barrientos is Postdoctoral Associate, Department of Statistical Science, Duke University, Durham, NC 27708 (E-mail: \href{mailto:andres.barrientos@duke.edu}{andres.barrientos@duke.edu}); and Jerome P. Reiter is Professor of Statistical Science, Duke University, Durham, NC 27708 (E-mail: \href{mailto:jreiter@duke.edu}{jreiter@duke.edu}). This research was supported by grants from the National Science Foundation (SES-11-31897) and the Alfred P. Sloan Foundation (G-2015-20166003). 
	}  }
\title{Simultaneous Edit and Imputation For Household Data with Structural Zeros}
\date{}

\begin{document}
\maketitle

\begin{abstract}
\noindent Multivariate categorical data nested within households often include reported values that fail edit constraints---for example, a participating household reports a child's age as older than his biological parent's age---as well as missing values.  Generally, agencies prefer datasets to be free from erroneous or missing values before analyzing them or disseminating them to secondary data users.  We present a model-based engine for editing and imputation of household data based on a Bayesian hierarchical model that includes (i) a nested data Dirichlet process mixture of products of multinomial distributions as the model for the true latent values of the data, truncated to allow only households that satisfy all edit constraints, (ii) a model for the location of  errors, and (iii) a reporting model for the observed responses in error. The approach propagates uncertainty due to unknown locations of errors and missing values, generates plausible datasets that satisfy all edit constraints, and can preserve multivariate relationships within and across individuals in the same household. We illustrate the approach using data from the 2012 American Community Survey. 
\end{abstract}


\section{Introduction}
In demographic surveys and population censuses, agencies often collect data on individuals nested within households.  Some of the variables correspond to household level characteristics, for example, whether or not the residents own the house, and other variables correspond to individual level characteristics, for example, the age of each resident. Typically, the reported data include erroneous values, i.e., combinations of variables that are inconsistent or theoretically impossible. For example, it should not be the case that a five year old child is married or that a parent is younger than her biological child.  Such erroneous values can arise due to data processing errors, e.g.,  when the age of an individual is erroneously recorded by the data collecting agency as 5 instead of 50, or respondent errors, e.g., when a household head responding to a survey accidentally selects the ``relationship to household head'' status of his/her biological parent as a child.

Agencies generally prefer not to analyze or disseminate data with overt errors.  The errors can affect inferences, potentially resulting in misleading conclusions.  When included in data releases, errors can undermine data users' confidence in the quality of the data. On the other hand, inferences based only on the subset of data without overt errors can be inefficient or even biased, depending on the reasons why values are in error \citep{Rubin1976,LittleRubin2002}.  It is therefore prudent for agencies to edit faulty data in hopes of improving quality before analysis or dissemination. 

When confronted with errors, ideally the agency can re-contact respondents to ascertain their true responses. However, this can be expensive and impractical to do for all respondents, especially in the context of censuses or large surveys.  
Many agencies therefore supplement re-contact operations with a process known as automatic edit-imputation. In the edit step, agencies specify an error localization process to determine a set of values that are in error for each record. This is usually done using a variant of the error localization suggested by \citet{FellegiHolt1976}, where the values are selected by minimizing the number of fields necessary to turn an erroneous record into a theoretically valid one \citep{Winkler1995, WinklerPetkunas1997}. In the imputation step, the values selected in the error localization are replaced with plausibly valid entries \citep{DeWaalCoutinho2005, DeWaalEtAl2011}. This is usually done by some form of hot deck imputation \citep{KaltonKasprzyk1986, AndridgeLittle2010}.


\citet{KimEtAl2015b, Manrique-VallierReiter2018} describe some shortcomings of edit-imputation approaches based on the Fellegi-Holt paradigm for non-nested data. In particular, they highlight two problems, namely that (i) the error localization process typically does not fully take advantage of  multivariate relationships in the data, and (ii) the selection of a single error localization coupled with a single imputation underestimates uncertainty. These shortcomings can be relevant in household data in  complicated ways. To illustrate, suppose the reported data include a household with a head, a spouse, and three biological children, two of whom are reported as age 6 and 8 and a third reported as age 30.  The reported age 30 exceeds the age of the household head.  Most likely, the reported 30 year old has at least one field in error; the person's age or relationship to household head is likely erroneous.  Agencies following the Fellegi-Holt paradigm would change one of these two fields based on some heuristic, e.g., change the variable that is more likely to have errors according to  experience in similar datasets. Now suppose the data also inform us that everyone except the reported 30 year child has the same race of Asian, and the 30 year old reports a race of black.  The data may well indicate that biologically-related families with two Asian parents, two Asian children, and one black child are highly unlikely.  Thus, it may be more plausible to leave age alone and change the relationship value to ``unrelated'' rather than leave relationship alone and change age.  Of course, it may well be that multiple fields are in error, including race.  In any case, we would like to incorporate the uncertainty in the error localization by averaging over plausible localizations and corrected values.


In this article, we present an edit-imputation approach intended to address these shortcomings.  Specifically, we follow the approach of 
\citet{KimEtAl2015b} and \citet{Manrique-VallierReiter2018} and handle the edit and imputation processes simultaneously with a Bayesian hierarchical model. The hierarchical model includes (i) a multivariate model for the true latent values of the data which has support only on theoretically possible households, (ii) a model for locations of errors given the latent true values, and (iii) a model for the reported values for fields in error.  For the multivariate model of the true values, we use the  
truncated nested data Dirichlet process mixture of products of multinomial distributions (NDPMPM) of  \citet{HuEtAl2018}.  This model includes household level and individual level variables, allows for within-household dependence, and puts zero probability on impossible combinations.  For the error location and reporting models, we adapt the measurement error model used for non-nested categorical data in \citet{Manrique-VallierReiter2018}.  We discuss alternative measurement error models in Section \ref{measurementerror} and Section \ref{discussion}.  We use a Markov chain Monte Carlo sampler to fit the hierarchical model, which generates plausible datasets without errors as byproducts.  These can be analyzed or disseminated as multiple imputations \citep{Rubin1987,Ghosh-DastidarSchafer2003}.  Our approach leverages the work of \citet{AkandeEtAl2017b}, who extended the NDPMPM to handle missing values. \citet{AkandeEtAl2017b} do not consider how to use the NDPMPM for edit and imputation of erroneous values, which is the primary  contribution of our work.

The remainder of this article is organized as follows. In Section \ref{EIHD}, we present the Bayesian edit-imputation model for household data, which we refer to as the EIHD model. 
In Section \ref{mcmc}, we describe the MCMC algorithm for fitting the EIHD. In Section \ref{simulations}, we report the results of simulation studies used to illustrate the performance of the EIHD.  The simulations are based on a subset of data from the 2012 American Community Survey (ACS) public use files.   In Section \ref{discussion}, we discuss findings, caveats and future work.  The EIHD is implemented in the R package, ``NestedCategBayesImpute,''  available on CRAN.  Source code is at \url{https://github.com/akandelanre/Edit-Imputation-for-Nested-Data/tree/master}.

\section{The EIHD Model} \label{EIHD}
In describing the EIHD, we use the notation of  \citet{HuEtAl2018} and \citet{Manrique-VallierReiter2018}. For the $i=1, \dots, n$ households in the data, let $n_i$ be the number of individuals in the household, so that there are  $\sum_{i=1}^n n_i = N$ individuals in the data. For $k = p+1, \ldots, p+q$, let $X_{ik}^1 \in \{1, \ldots, d_k\}$ be the true value of household level variable $k$ for household $i$, which is assumed to be identical for all $n_i$ individuals in household $i$.  Similarly, for $k = 1, \ldots, p$ and $j = 1, \ldots, n_i$, for each household let $X_{ijk}^1 \in \{1, \ldots, d_k\}$ be the true value of individual level variable $k$ for person $j$ in household $i$. We associate each household $i$ with $\textbf{X}_i^1 = (X_{i(p+1)}^1, \dots, X_{i(p+q)}^1, X_{i11}^1, \dots, X_{in_ip}^1)$, which includes all household level and individual level variables for the $n_i$ individuals in the household. 

Let $\mathcal{X}^1 = (\textbf{X}_{1}^1, \ldots, \textbf{X}_{n}^1)$. We do not observe $\mathcal{X}^1$; instead, we observe the reported data $\mathcal{Y} = (\textbf{Y}_{1}, \ldots, \textbf{Y}_{n})$. Each $\textbf{Y}_{i}$ is a potentially contaminated version of its corresponding $\textbf{X}_{i}^1$. We assume that each $\textbf{Y}_{i}$ is generated conditional on $\textbf{X}_{i}^1$ through a measurement error model, with density $f_Y(\textbf{Y}| \textbf{X}^1, \theta_{\textbf{Y}} )$.
Whenever it is the case that $\textbf{Y}_i \neq \textbf{X}_{i}^1$, we say that the observed data for the $i$-th household contains errors. 

Let $\mathcal{H}$ be the set of all household sizes that are possible in the population. Let $\mathcal{C} = \bigcup_{h \in \mathcal{H}} \mathcal{C}_h$, where each $\mathcal{C}_h$ represents the set of all combinations of individual-level and household-level variables for households of size $h$, that is, $\mathcal{C}_h = \prod_{k=p+1}^{p+q} \{1, \ldots, d_k\} \prod_{j=1}^{h} \prod_{k=1}^{p} \{1, \ldots, d_k\}$. Let $\mathcal{S}_h \subset \mathcal{C}_h$ represent the set of impossible combinations defined by a set of edit rules for households of size $h$, i.e., the structural zeros \citep{BishopEtAl1975}, for which $\Pr(\textbf{X}_{i}^1 \in \mathcal{S}_h) = 0$. These can include combinations of variables for any individual, e.g., a five year old person cannot be a parent, or across individuals in the same household, e.g., a person cannot be younger than his biological child. Let $\mathcal{S} = \bigcup_{h \in \mathcal{H}} \mathcal{S}_h$. 

Although true responses are such that $\textbf{X}_{i}^1 \notin \mathcal{S}_{n_i}$, reported responses potentially can violate the edit rules, i.e., $\Pr(\textbf{Y}_{i} \in \mathcal{S}_{n_i}) > 0$. Whenever $\textbf{Y}_{i} \in \mathcal{S}_{n_i}$, we know for sure that $\textbf{Y}_{i}$ contains errors. We refer to such errors as detectable.  Though the errors may be detectable, the exact location of the errors is usually unknown. For example, suppose a household contains a male household head who is 35 years old and his biological child who is 60 years old. Certainly, this household contains errors but we cannot say for sure whether the error is in the ages of at least one of the individuals, the relationship between them (since the 60 year old could in fact be a parent instead of a biological child), or both.

It may be possible that $\textbf{Y}_{i} \notin \mathcal{S}_{n_i}$ while $\textbf{Y}_{i} \neq \textbf{X}_{i}^1$.   We refer to such errors as undetectable. In this article, we make the simplifying assumption that the only errors in the data are detectable ones. While this assumption can be viewed as somewhat unrealistic, it is consistent with the practice of most statistical agencies that use automatic edit-imputation algorithms, including Fellegi-Holt systems  \citep{DeWaalEtAl2011, KimEtAl2015b}.  It stems from a philosophy that agencies should change as few respondents' reported values as possible. We describe how to relax this assumption at the end of  Section \ref{measurementerror}. 

Finally, we assume that each unobserved $\textbf{X}_{i}^1$ is stochastically generated from a common data generating process with density $f_{X}(\textbf{X}^1 | \theta_{\textbf{X}})$ and support restricted to $\textbf{X}^1 \in \mathcal{C} - \mathcal{S}$, so that the realized values for $\mathcal{X}^1$ must satisfy the structural zero rules. Under this setup, the objective is to estimate the joint distribution of the underlying true data $\mathcal{X}^1$ and the erroneous data $\mathcal{Y}$, and obtain posterior predictive samples of $\mathcal{X}^1$ from it.

\subsection{True response model} \label{truerespmodel}
In theory, $f_{X}(\textbf{X}^1 | \theta_{\textbf{X}})$ can be any multivariate categorical data model that adequately describes the joint distribution of all the variables, has support restricted to $\mathcal{C} - \mathcal{S}$, and captures the relevant structure in $\mathcal{X}^1$. For household data, the truncated NDPMPM model of \citet{HuEtAl2018} has those properties. 
In this section, we briefly review the 
 NDPMPM model. We refer readers to \citet{HuEtAl2018} for a detailed development of the model.

Each household $i$ belongs to one of $F$ household level classes representing latent household types. For $i = 1, \dots, n$, let  
$G_i^1 \in (1, \dots, F)$ indicate the household level latent class for household $i$. Let $\pi_g = \Pr(G_i^1 = g)$ be the probability that household $i$ belongs to class $g$. 
For any $k \in \{p+1,\ldots,p+q\}$ and any $c \in \{1,\ldots,d_k\}$, let $\lambda_{gc}^{(k)} = \Pr(X_{ik}^1 = c | G_i^1 = g)$ for any class $g$, where $\lambda_{gc}^{(k)}$ is the same value for every household in class $g$.  Let $\pi = \{\pi_1, \ldots \pi_F\}$, $\lambda = \{\lambda^{(k)}_{gc}: c = 1, \ldots, d_k; k = p+1, \ldots, p+q; g = 1, \ldots F\}$, and $\textbf{G}^1 = \{G_i^1: i=1,\ldots,n\}$.

Within each household class, each individual belongs to one of $S$ individual level latent classes. 
For $i=1, \dots, n$ and $j=1, \dots, n_i$, let $M_{ij}^1$ represent the individual level latent class of individual $j$ in household $i$. Let $\omega_{gm} = \Pr(M_{ij}^1 = m | G_i^1 = g)$ be the probability that individual $j$ in household $i$ belongs to individual-level class $m$ nested within household-level class $g$.
For any $k \in \{1,\ldots,p\}$ and any $c \in \{1,\ldots,d_k\}$, let $\phi_{gmc}^{(k)} = \Pr(X_{ijk}^1 = c | (G_i^1, M_{ij}^1) = (g,m))$ for the class pair $(g,m)$, where $\phi_{gmc}^{(k)}$ is the same value for every individual in the class pair $(g,m)$. Let $\omega = \{\omega_{gm}: g = 1, \ldots F; m=1, \ldots, S \}$, $\phi = \{\phi^{(k)}_{gmc}: c = 1, \ldots, d_k; k = 1, \ldots, p; m=1, \ldots, S ; g = 1, \ldots F\}$, and $\textbf{M}^1 = \{M_{ij}^1: i=1,\ldots,n; j=1,\ldots,n_i\}$. 

To introduce the model, we first present the version of the NDPMPM model appropriate for data without 
structural zeros.  This generative model can be written as
\begin{align}
\phantom{X_{ijk}^1 | G_i^1, M_{ij}^1, \phi, n_i}
&\begin{aligned} \label{ModelSpecification1}
\mathllap{X_{ik}^1 | G_i^1, \lambda} & \sim \textrm{Discrete}(\lambda^{(k)}_{G_i^11}, \ldots, \lambda^{(k)}_{G_i^1d_k}) \ \ \textrm{for} \ k = p + 1, \ldots, p + q 
\end{aligned}\\
&\begin{aligned} \label{ModelSpecification2}
\mathllap{X_{ijk}^1 | G_i^1, M_{ij}^1, \phi, n_i}  & \sim \textrm{Discrete}(\phi^{(k)}_{G_i^1M_{ij}^11}, \ldots, \phi^{(k)}_{G_i^1M_{ij}^1d_k})  \ \ \textrm{for} \ j = 1, \ldots, n_i; \ \ k = 1, \ldots, p
\end{aligned}\\
&\begin{aligned} \label{ModelSpecification3}
\mathllap{G_i^1 | \pi} & \sim \textrm{Discrete}(\pi_1, \ldots, \pi_F)
\end{aligned}\\
&\begin{aligned} \label{ModelSpecification4}
\mathllap{M_{ij}^1 | G_i^1, \omega, n_i} & \sim \textrm{Discrete}(\omega_{G_i^11}, \ldots, \omega_{G_i^1S})  \ \ \textrm{for} \   j = 1, \ldots, n_i
\end{aligned}
\end{align}
where $i = 1, \ldots, n$. Here, (\ref{ModelSpecification2}) and (\ref{ModelSpecification4}) are conditioned on $n_i$ so that the model can be interpreted as a generative model for households. That is, the household size is first sampled from (\ref{ModelSpecification1}), and once the size is known, the characteristics of the household's individuals are sampled from (\ref{ModelSpecification2}) and (\ref{ModelSpecification4}). Without loss of generality, we set $n_i$ to be the first household variable, that is, $X_{i(p+1)}^1 = n_i$. The distributions in (\ref{ModelSpecification2}) and (\ref{ModelSpecification4}) do not depend on $n_i$ other than to fix the number of people in the household; that is, within any $G_i^1$, the distributions of all parameters do not depend on $n_i$.

For prior distributions, we follow the recommendations of \citet{HuEtAl2018}. We use independent uniform Dirichlet distributions as priors for $\lambda$ and $\phi$, and the truncated stick-breaking representation of the Dirichlet process as priors for $\pi$ and $\omega$ \citep{Sethuraman1994, DunsonXing2009, SiReiter2013, Manrique-VallierReiter2014b}. We have
\begin{align}
\phantom{\omega_{gm} }
&\begin{aligned} \label{DirichletPrior}
\mathllap{\lambda_g^{(k)}} & \sim \textrm{Dirichlet}(1,\ldots, 1); \ \ \ \phi^{(k)}_{gm} \sim \textrm{Dirichlet}(1,\ldots, 1) \\ 
\end{aligned}\\
&\begin{aligned}
\mathllap{\pi_g } & = u_g \prod_{f < g} (1 -  u_f) \ \textrm{for} \ g = 1, \ldots F; \ \ \ u_1, \ldots, u_{F-1} \overset{iid}{\sim} \textrm{Beta}(1,\alpha), \ u_F = 1 \\
\end{aligned}\\
&\begin{aligned}
\mathllap{\omega_{gm} } & = v_{gm} \prod_{s < m} (1 -  v_{gs}) \ \textrm{for} \ m = 1, \ldots S; \ \ \ v_{g1}, \ldots, v_{gS-1} \overset{iid}{\sim} \textrm{Beta}(1,\beta), \ v_{gS} = 1 \\
\end{aligned}\\
&\begin{aligned} \label{GammaPrior}
\mathllap{\alpha } & \sim \textrm{Gamma}(0.25, 0.25); \ \ \ \beta \sim \textrm{Gamma}(0.25, 0.25) \\ 
\end{aligned}
\end{align}
where $\lambda_g^{(k)} = (\lambda^{(k)}_{g1}, \ldots, \lambda^{(k)}_{gd_k})$ and $\phi^{(k)}_{gm} = (\phi^{(k)}_{gm1}, \ldots, \phi^{(k)}_{gmd_k})$. We set the parameters for the Dirichlet distributions in (\ref{DirichletPrior}) to $\mathbf{1}_{d_k}$ (a $d_k$-dimensional vector of ones) and the parameters for the Gamma distributions in (\ref{GammaPrior}) to $0.25$ to represent vague prior specifications. For further discussions on prior specifications, see \citet{HuEtAl2018}.

The model in (\ref{ModelSpecification1})---(\ref{ModelSpecification4}) does not incorporate structural zeros.  
To do so, \citet{HuEtAl2018} truncate the model to assign zero probability to $\mathcal{S}$. The likelihood of the truncated NDPMPM,  $L(\theta_{\textbf{X}} | \mathcal{X}^1)$ then takes the form
\begin{equation} \label{StrucZeroLikelihood}
L(\theta_{\textbf{X}} | \mathcal{X}^1) \ \propto \ \prod_{i=1}^n \sum_{h \in \mathcal{H}} \left( \mathds{1}\{n_i = h \} \mathds{1}\{\textbf{X}_i^1 \notin \mathcal{S}_h \} \sum_{g=1}^F \pi_g \left[ \prod^{p+q}_{k = p+1} \lambda^{(k)}_{gX^1_{ik}} \prod^{h}_{j=1} \sum_{m=1}^S \omega_{gm}\prod^p_{k = 1} \phi^{(k)}_{gmX^1_{ijk}} \right] \right)
\end{equation}
where $\theta_{\textbf{X}}$ includes all the parameters of the untruncated NDPMPM and $\mathds{1}\{.\}$ equals one when the condition inside the $\{\}$ is true and equals zero otherwise. 

\citet{HuEtAl2018} use a data augmentation strategy  to estimate the posterior distribution based on the truncated likelihood in (\ref{StrucZeroLikelihood}). Let random variables that have support only on $\mathcal{S}$ be indexed with the superscript ``$0$'' and those with unrestricted support on $\mathcal{C}$ have no superscript. For example, $\textbf{X}_i^1$ is a random variable with support on $\mathcal{C} - \mathcal{S}$ whereas $\textbf{X}_i^0$ has support on $\mathcal{S}$ and $\textbf{X}_{i}$ has unrestricted support on $\mathcal{C}$. \citet{HuEtAl2018} view $\mathcal{X}^1$ as a subset from a hypothetical sample $\mathcal{X}$ of $(n + n_0)$ households directly generated from the untruncated NDPMPM, where $n_0$ is the number of households in the unobserved data, $\mathcal{X}^0 = \mathcal{X} - \mathcal{X}^1$, that fail the structural zeros. Let $n_i^0$ be the number of individuals in each $i=1, \dots, n_0$ households in $\mathcal{X}^0$. We do not observe the household level and individual level class indicators $\textbf{G}^0 = \{G_i^0: i=1,\ldots,n_0\}$ and $\textbf{M}^0 = \{M_{ij}^0: i=1,\ldots,n_0; j=1,\ldots,n_i^0\}$. \citet{HuEtAl2018} sample the augmented data $(\textbf{G}^0,\textbf{M}^0,\mathcal{X}^0,n_0)$ using a rejection sampler.  They show that, assuming a uniform prior for $(n + n_0)$, the marginal posterior of the model parameters after integrating out the augmented data directly matches the truncated posterior distribution of $\theta_{\textbf{X}}$ based on (\ref{StrucZeroLikelihood}). Details and proof that the augmentation scheme converges to the desired posterior distribution are in \citet{HuEtAl2018}.

\subsection{Measurement error model}\label{measurementerror}
For the measurement error model, we introduce a series of binary indicator variables to help with model specification. 
Let $Z_i = 1$ if household $i$ has an error and $Z_i=0$ otherwise.  Let $E_{ik} = 1$ if household level variable $k$ is in error for household $i$, and $E_{ik} = 0$ otherwise.  Let $E_{ijk} = 1$ if individual level variable $k$ is in error for person $j$ in household $i$, and $E_{ijk} = 0$ otherwise.   By design, $Z_i = 0$ implies that $E_{ik} = E_{ijk} = 0$ for all $j$ and $k$ corresponding to household $i$, whereas $Z_i = 1$ implies that at least one of the $E_{ik}$ and $E_{ijk}$ corresponding to household $i$ equal one. Since we assume no undetectable errors,
each $Z_i$ is observed rather than latent. This saves computational time, since whenever $\textbf{Y}_{i} \notin \mathcal{S}_{n_i}$, we set $\textbf{X}_{i}^1 = \textbf{Y}_{i}$ and do not need to sample a new plausible value for $\textbf{X}_{i}^1$. Finally, let $\mathbf{E} = (\textbf{E}_1, \ldots, \textbf{E}_n)$, where each $\textbf{E}_i = (E_{1(p+1)}, \dots, E_{i(p+q)}, E_{i11}, \dots, E_{in_ip})$.

We formulate the measurement error model as two sub-models, namely (i) a reporting model for $\textbf{Y}_{i}$ conditional on $\textbf{X}_{i}^1$ and $\mathbf{E}_i$, and (ii) an error location model for $\mathbf{E}_i$ conditional on the $Z_i$.  For the reporting model for any $Y_{ijk}$ or $Y_{ik}$, we have 
\begin{align}
\phantom{Y_{ijk} | X_{ijk}^1 = c, E_{ijk} = e}
&\begin{aligned} \label{ReportingModel1}
\mathllap{Y_{ijk} | X_{ijk}^1 = c, E_{ijk} = e} & \sim
\begin{cases}
\delta_{X_{ijk}^1} & \textrm{if} \ e = 0 \\
\textrm{Discrete}(\{1, \ldots, d_k\} \backslash \{c\}; \{q_k, \ldots, q_k\} ) & \textrm{if} \ e = 1 \\
\end{cases} \\
& \forall \ i, j; k \in \{1, \ldots, p\}
\end{aligned}\\
&\begin{aligned} \label{ReportingModel2}
\mathllap{Y_{ik} | X_{ik}^1 = c, E_{ik} = e}  & \sim
\begin{cases}
\delta_{X_{ik}^1} & \textrm{if} \ e = 0 \\
\textrm{Discrete}(\{1, \ldots, d_k\} \backslash \{c\}; \{q_k, \ldots, q_k\} ) & \textrm{if} \ e = 1 \\
\end{cases} \\
& \forall \ i; k \in \{p+1, \ldots, p+q\}
\end{aligned}
\end{align}
where $q_k = 1/(d_k - 1)$.  This model assumes that each $Y_{ijk}$ (and $Y_{ik}$)  in error is generated uniformly from the set of all possible values for variable $k$, excluding the true value $X_{ijk}^1$ (and $X_{ik}^1$).  It implies that when people make reporting errors, they do so completely randomly and independently across variables.  One could replace these assumptions with informative models, should such information be available from other data sources or previous experience.  For example, one could specify reporting models that put higher probability on categories adjacent to the true $X_{ijk}^1$, reflecting response errors where people mistakenly select nearby categories on a survey form. 
	
For the error location model for any $E_{ijk}$ or $E_{ik}$, we have 
\begin{align}
\phantom{E_{ijk} | Z_i = z, \epsilon_k}
&\begin{aligned} \label{ErrorModel1}
\mathllap{E_{ijk} | Z_i = z, \epsilon_k} & \sim
\begin{cases}
\delta_{0} & \textrm{if} \ z = 0 \\
\textrm{Bernoulli}(\epsilon_k) & \textrm{if} \ z = 1 \\
\end{cases} \ \ \ \forall \ i, j; k \in \{1, \ldots, p\} \\
\end{aligned}\\
&\begin{aligned} \label{ErrorModel2}
\mathllap{E_{ik} | Z_i = z, \epsilon_k}  & \sim
\begin{cases}
\delta_{0} & \textrm{if} \ z = 0 \\
\textrm{Bernoulli}(\epsilon_k) & \textrm{if} \ z = 1 \\
\end{cases} \ \ \ \forall \ i; k \in \{p+1, \ldots, p+q\} \\
\end{aligned}
\end{align}
This model assumes that the error indicators are independent across variables, which again accords with people making errors at random although possibly with different rates for different variables.  One could replace these assumptions with models conditional on the true values, e.g., people who are older are more likely to make errors on certain variables.  When conditioning on true values that are latent, this creates a nonignorable faulty data mechanism.

We complete the specification with prior distributions for $\boldsymbol{\epsilon} = \{\epsilon_k: k=1,\ldots,p+q\}$. In the empirical study with the ACS data, we use conjugate beta priors for each $\epsilon_k$, primarily for computational convenience. We have 
\begin{equation}
\epsilon_k \sim \textrm{Beta}(a_{\epsilon_k}, b_{\epsilon_k}) \ \ \forall \ k = 1, \ldots, p, p+1, \ldots, p+q.
\end{equation}
We set  $a_{\epsilon_k} = b_{\epsilon_k} = 1$ for each $k \in \{1, \ldots, p, p+1, \ldots, p+q\}$ to represent complete ignorance about the true error rates. In applied contexts, we suggest setting each $(a_{\epsilon_k}, b_{\epsilon_k})$ to reflect prior beliefs on the error rates whenever reasonable prior information is available. 

The measurement error model can be  extended to allow for undetectable errors by letting $Z_i$ be latent for $\mathbf{Y}_{i} \notin \mathcal{S}$. We continue to let $Z_i = 1$ for cases where $\mathbf{Y}_{i} \in \mathcal{S}$.  When $\mathbf{Y}_{i} \notin \mathcal{S}$ for example, we can let $Z_i | \rho \sim \textrm{Bernoulli}(\rho)$, with $\rho \sim \textrm{Beta}(a_{\rho}, b_{\rho})$ reflecting prior beliefs about the fraction of cases with errors. We leave investigation of this model for future research. 

The model also can handle missing values simultaneously with faulty data. One sets $Z_i = 1$ for households that contain at least one missing entry and $E_{ik} = E_{ijk} = 1$ for all variables that have missing values, forcing $\textbf{X}_{i}^1$ to be imputed for those households. This presumes that (i) the values are missing at random, and (ii) the same measurement error and true response models apply for the households with error and the households with missing data.

\section{MCMC Estimation} \label{mcmc}
We use a Gibbs sampler to estimate the posterior distribution of the parameters in the EIHD model.  Given the data $\mathcal{Y}$ and a draw of $(\textbf{G}^1,\textbf{G}^0,\textbf{M}^1,\textbf{M}^0,\mathcal{X}^0,\theta_{\textbf{X}},n_0)$, we update $(\mathcal{X}^1, \mathbf{E}, \boldsymbol{\epsilon})$. We outline these updates in Section \ref{Gibbs1}. We then update $(\textbf{G}^1,\textbf{G}^0,\textbf{M}^1,\textbf{M}^0,\mathcal{X}^0,\theta_{\textbf{X}},n_0)$ given a draw of $\mathcal{X}$. We outline these updates in Section \ref{Gibbs2}.

Upon convergence of the Gibbs sampler, analysts can obtain posterior inferences for parameters of interest or treat the posterior samples of $\mathcal{X}^1$ as multiply imputed datasets \citep{Rubin1987}. For the latter, analysts can select a modest number $L$ of datasets (usually, $L \geq 5$), reasonably spaced so that they are approximately independent.

\subsection{Sampling $(\mathcal{X}^1, \mathbf{E}, \boldsymbol{\epsilon})$} \label{Gibbs1}

Sampling $\epsilon_k$ for each variable $k$ is straightforward since error rates are independent across variables. We use the following step in the sampler.
\begin{enumerate}
	\item[S1.] For $k = 1, \ldots, p, p+1, \ldots, p+q$, sample $\epsilon_k | \ldots \sim \textrm{Beta}(a_{\epsilon_k} + a^\star_{\epsilon_k}, b_{\epsilon_k} + b^\star_{\epsilon_k})$, where
	\begin{equation*}
	\begin{split}
	a^\star_{\epsilon_k} & = \sum\limits_{i| Z_i = 1}^{} \mathds{1}(E_{ik} = 1), \ \ b^\star_{\epsilon_k} = \sum\limits_{i| Z_i = 1}^{} \mathds{1}(E_{ik} = 0) \ \ \textrm{if} \ k \in \{p+1, \ldots, p+q\}; \ \ \textrm{and} \\
	a^\star_{\epsilon_k} & = \sum\limits_{i| Z_i = 1}^{} \sum\limits_{j=1}^{n_i} \mathds{1}(E_{ijk} = 1),\ \ b^\star_{\epsilon_k} = \sum\limits_{i| Z_i = 1}^{} \sum\limits_{j=1}^{n_i} \mathds{1}(E_{ijk} = 0) \ \ \textrm{if} \ k \in \{1, \ldots, p\}.
	\end{split}
	\end{equation*}
\end{enumerate}	

Sampling $(\mathcal{X}^1, \mathbf{E})$ is more involved. Since each $\mathbf{E}_i$ is completely determined by $\mathbf{X}_i^1$ and $\mathbf{Y}_i$, we cannot form independent Gibbs steps for each using the full conditionals $\Pr(\textbf{X}_i^1 | \ldots)$ and $\Pr(\textbf{E}_i | \ldots)$. Instead, we sample directly from $\Pr(\textbf{X}_i^1, \textbf{E}_i| \ldots)$ using the factorization 
$$\Pr(\textbf{X}_i^1, \textbf{E}_i| \ldots) = \Pr(\textbf{X}_i^1| \ldots, -\{\textbf{E}_i\} ) \times \Pr(\textbf{E}_i| \ldots)$$
where $\Pr(\textbf{X}_i^1| \ldots, -\{\textbf{E}_i\} )$ is the conditional pmf of $\mathbf{X}_i^1$ given all other random variables in the model except $\textbf{E}_i$. We therefore sample $(\mathcal{X}^1, \mathbf{E})$ using the following steps.
\begin{enumerate}
	\item[S2.] For $i = 1, \ldots, n$, set $\mathbf{X}_i^1 = \mathbf{Y}_i$ if $Z_i = 0$. If $Z_i = 1$, sample $\mathbf{X}_i^1$ from
	\begin{equation*}
	 \Pr(\textbf{X}_i^1| \ldots, -\{\textbf{E}_i\} ) \ \propto \ \mathds{1}\{\textbf{X}_i^1 \notin \mathcal{S}_h \} \pi_{G_i^1} \left[\prod^{p+q}_{k = p+1} \lambda^{(k)(\star)}_{G_i^1X_{ik}^1} \left(\prod^{n_i}_{j=1} \omega_{G_i^1M_{ij}^1}\prod^p_{k = 1} \phi^{(k)(\star)}_{G_i^1M_{ij}^1X_{ijk}^1} \right) \right], \ \ \textrm{where}
	\end{equation*}
	\begin{equation*}
	\begin{split}
	\lambda^{(k)(\star)}_{G_i^1X^1_{ik}} & = \lambda^{(k)}_{G_i^1X^1_{ik}} (1 - \epsilon_k)^{\mathds{1}\{Y_{ik} = X^1_{ik}\}} (q_k \epsilon_k)^{\mathds{1}\{Y_{ik} \neq X^1_{ik}\}}, \ \ \textrm{and} \\
	\phi^{(k)(\star)}_{G_i^1M_{ij}^1X^1_{ijk}} & = \phi^{(k)}_{G_i^1M_{ij}^1X^1_{ijk}} (1 - \epsilon_k)^{\mathds{1}\{Y_{ijk} = X^1_{ijk}\}} (q_k \epsilon_k)^{\mathds{1}\{Y_{ijk} \neq X^1_{ijk}\}}.
	\end{split}
	\end{equation*}
	Sampling from this conditional distribution is nontrivial because of the dependence among variables induced by the structural zero rules in each $\mathcal{S}_h$. However, we can generate the samples through the following rejection sampling scheme.
	\begin{enumerate}
		\item Set $X_{i(p+1)}^1 = n_i$. For the remaining household level variables $k \in \{p+2, \ldots, p+q \}$, sample $X_{ik}^{1} \sim \textrm{Discrete}(\lambda^{(k)(\star)}_{G_i^11}, \ldots, \lambda^{(k)(\star)}_{G_i^1d_k})$.
		
		\item Sample $X_{ijk}^{1} \sim \textrm{Discrete}(\phi^{(k)(\star)}_{G_i^1M_{ij}^11}, \ldots, \phi^{(k)(\star)}_{G_i^1M_{ij}^1d_k})$ for each $k \in \{1, \ldots, p\}$ and $j = 1, \ldots, n_i$.
		
		\item Set the sampled household level and individual level values to $\textbf{X}_i^{1\star} $. 
		
		\item If $\textbf{X}_i^{1\star} \notin \mathcal{S}_h$, set $\textbf{X}_i^{1} = \textbf{X}_i^{1\star}$, otherwise, return to step (a).
	\end{enumerate}
We then	set $\mathcal{X}^1 = (\textbf{X}^1_{1}, \ldots, \textbf{X}^1_{n})$.  We then set $E_i$ deterministically as follows.
	
	\item[S3.] For $i = 1, \ldots, n$ and $k \in \{p+1, \ldots, p+q \}$, set $E_{ik} = 1$ if  $X^1_{ik} \neq Y_{ik}$ and $E_{ik} = 0$ otherwise. Similarly, for $i = 1, \ldots, n$, $j = 1, \ldots, n_i$ and $k \in \{1, \ldots, p\}$, set $E_{ijk} = 1$ if  $X^1_{ijk} \neq Y_{ijk}$ and $E_{ijk} = 0$ otherwise.
\end{enumerate}

\subsection{Sampling $(\textbf{G}^1,\textbf{G}^0,\textbf{M}^1,\textbf{M}^0,\mathcal{X}^0,\theta_{\textbf{X}},n_0)$ } \label{Gibbs2}

To sample $(\textbf{G}^1,\textbf{G}^0,\textbf{M}^1,\textbf{M}^0,\mathcal{X}^0,\theta_{\textbf{X}},n)$, we use the version of the Gibbs sampler for the NDPMPM proposed in \citet{AkandeEtAl2017b} for multiple imputation of missing nested categorical data (absent any erroneous values), which they call the cap-and-weight approach.
Let $n_{h}^0$ be the number of households of size $h$ in $\mathcal{X}^0$ and $n_{h}^1$ be the number of households of size $h$ in $\mathcal{X}^1$, so that $n_0 = \sum_h n_{h}^0$ and $n = \sum_h n_{h}^1$. \citet{AkandeEtAl2017b} put an upper bound on the number of cases in  $\mathcal{X}^0$ sampled at each MCMC iteration by sampling $\lceil n_{h}^0 \times \psi_h \rceil$ impossible households for each $h \in \mathcal{H}$ (instead of $n_{h}^0$ households) for some $\psi_h$ such that $1/\psi_h$ is a positive integer. This approach speeds up the sampler by approximating the likelihood of the full unobserved data with a pseudo likelihood using weights (the $1/\psi_h$). Setting each $\psi_h = 1$ results in the original rejection sampler in \citet{HuEtAl2018}, so that the cap-and-weight approach should in practice provide similar results to the original rejection sampler when all $\psi_h$ are near $1$. 

At each MCMC iteration, we do the following steps.
\begin{enumerate}
	\item[S4.] Set $\mathcal{X}^0 = \textbf{G}^0 = \textbf{M}^0 = \emptyset$. For each $h \in \mathcal{H}$, repeat:
	
	\begin{enumerate}
		\item Set $t_0 = 0$ and $t_1 = 0$.
		
		\item Sample $G_i^0 \in \{1, \ldots, F \} \sim \textrm{Discrete}(\pi_1^{\star\star}, \ldots, \pi_F^{\star\star})$ where $\pi_g^{\star\star} \  \propto \ \lambda^{(k)}_{gh} \pi_g $ and $k$ is the index for the household-level variable ``household size''. 
		
		\item For $j = 1, \ldots, h$, sample $M^0_{ij} \in \{1, \ldots, S\} \sim \textrm{Discrete}(\omega_{G^0_i1}, \ldots, \omega_{G^0_iS})$.
		
		\item Set $X^0_{ik} = h$, where $X^0_{ik}$ corresponds to the variable for household size. Sample the remaining household level and individual level values using the likelihoods in (\ref{ModelSpecification1}) and (\ref{ModelSpecification2}). Set the household's simulated value to $\textbf{X}^0_i$.
		
		\item If $\textbf{X}^0_i \in \mathcal{S}_h$, let $t_0 = t_0 + 1$, $\mathcal{X}^0 = \mathcal{X}^0 \cup \textbf{X}^0_i$, $\textbf{G}^0 = \textbf{G}^0 \cup G^0_i$ and $\textbf{M}^0 = \textbf{M}^0 \cup \{M_{i1}^0, \ldots, M_{ih}^0 \}$. Otherwise set $t_1 = t_1 + 1$. \label{StepF}
		
		\item If $t_1 < \lceil n_{h}^1 \times \psi_h \rceil$, return to step (b). Otherwise, set $n_{h}^0 = t_0$. . \label{StepG}
	\end{enumerate}

	\item[S5.] For each household $i$ in $\mathcal{X}^1$, where $i = 1, \ldots, n$, 
	\begin{enumerate}
		\item Sample $G_i^1 \in \{1, \ldots, F \} \sim \textrm{Discrete}(\pi_1^\star, \ldots, \pi_F^\star)$, where 
		$$\pi_g^\star = \Pr(G_i^1 = g | \ldots ) = \dfrac{\pi_g \left[\prod\limits^q_{k=p+1} \lambda^{(k)}_{gX^1_{ik}} \left(\prod\limits^{n_i}_{j=1} \sum\limits^S_{m=1}\omega_{gm}\prod\limits^p_{k=1} \phi^{(k)}_{gmX^1_{ijk}} \right) \right] }{\sum\limits^F_{f=1} \pi_f \left[\prod\limits^q_{k=p+1} \lambda^{(k)}_{fX^1_{ik}} \left(\prod\limits^{n_i}_{j=1} \sum\limits^S_{m=1}\omega_{gm}\prod\limits^p_{k=1} \phi^{(k)}_{fmX^1_{ijk}} \right) \right] }$$
		for $g = 1, \ldots, F$.
		
		\item Sample $M^1_{ij} \in \{1, \ldots, S\} \sim \textrm{Discrete}(\omega_{G_i^11}^\star, \ldots, \omega_{G_i^1S}^\star)$ for each $j = 1, \ldots, n_i$, where 
		$$\omega_{G_i^1m}^\star = \Pr(M^1_{ij} = m | \ldots ) = \dfrac{\omega_{G_i^1m}\prod\limits^p_{k=1} \phi^{(k)}_{G_i^1mX^1_{ijk}}  }{ \sum\limits^S_{s=1}\omega_{G_i^1s}\prod\limits^p_{k=1} \phi^{(k)}_{G^1_isX^1_{ijk}}} \ \ \textrm{for} \ \ m = 1, \ldots, S.$$
	\end{enumerate}

	\item[S6.] For $g = 1, \ldots, F$, set $\pi_g = u_g \prod_{f<g} (1 - u_f)$, where 
	\begin{equation*}
	\begin{split}
	u_1,\ldots,u_{F-1} | \ldots \ & \overset{iid}{\sim} \textrm{Beta} \left(1 + U_g, \alpha + \sum^F_{f=g+1} U_f  \right), \ \ u_F = 1 \\
	\textrm{and} \ \ U_g & = \sum^{n}_{i=1} \mathds{1}(G^1_i = g) + \sum_{h \in \mathcal{H}} \dfrac{1}{\psi_h} \sum\limits_{i | n^0_i = h} \mathds{1}(G_i^0 = g)
	\end{split}
	\end{equation*}
	
	\item[S7.] For $g = 1, \ldots, F$ and $m = 1, \ldots, S$, set $\omega_{gm} = v_{gm} \prod_{s<m} (1 - v_{gs})$, where
	\begin{equation*}
	\begin{split}
	v_{gm},\ldots,v_{gS-1} | \ldots \ & \overset{iid}{\sim} \textrm{Beta} \left(1 + V_{gm}, \beta + \sum^S_{s=m+1} V_{gs}  \right), \ \ v_{gM} = 1 \\
	\textrm{and} \ \ V_{gm} & = \sum^{n}_{i=1} \mathds{1}(G^1_i = g, M^1_{ij} = m) + \sum_{h \in \mathcal{H}} \dfrac{1}{\psi_h} \sum\limits_{i | n_i^0 = h} \mathds{1}(G_i^0 = g, M_{ij}^0 = m) 
	\end{split}
	\end{equation*}
	
	\item[S8.] For $g = 1, \ldots, F$ and $k = p+1, \ldots, q$, sample 
	\begin{equation*}
	\begin{split}
	\lambda_g^{(k)} | \ldots & \sim \textrm{Dirichlet}\left(1 + \eta^{(k)}_{g1}, \ldots, 1 + \eta^{(k)}_{gd_k} \right) \\
	\textrm{where} \ \ \eta^{(k)}_{gc} & = \sum^{n}_{i|G^1_i = g} \mathds{1}(X^1_{ik} = c) + \sum_{h \in \mathcal{H}} \dfrac{1}{\psi_h} \sum\limits_{i \big| \substack{n_i^0 = h, \\ G_i^0 = g}} \mathds{1}(X_{ik}^0 = c)
	\end{split}
	\end{equation*}
	
	\item[S9.] For $g = 1, \ldots, F$, $m = 1, \ldots, S$ and $k = 1, \ldots, p$, sample 
	\begin{equation*}
	\begin{split}
	\phi_{gm}^{(k)} | \ldots & \sim \textrm{Dirichlet}\left(1 + \nu^{(k)}_{gm1}, \ldots, 1 + \nu^{(k)}_{gmd_k} \right) \\
	\textrm{where} \ \ \nu^{(k)}_{gmc} & = \sum^{n}_{i \big| \substack{G^1_i = g, \\  M^1_{ij} = m}} \mathds{1}(X_{ijk}^1 = c) + \sum_{h \in \mathcal{H}} \dfrac{1}{\psi_h} \sum\limits_{i \big| \substack{n_i^0 = h, \\ G_i^0 = g, \\  M^0_{ij} = m}} \mathds{1}(X_{ijk}^0 = c)
	\end{split}
	\end{equation*}
	
	\item[S10.] Sample
	$$ \alpha  | \ldots \sim \textrm{Gamma}\left(a_\alpha + F - 1, b_\alpha - \sum^{F-1}_{g=1} \textrm{log}(1-u_g) \right).$$
	
	\item[S11.] Sample
	$$ \beta  | \ldots \sim \textrm{Gamma}\left(a_\beta + F \times (S - 1), b_\beta - \sum^{S-1}_{m=1} \sum^{F}_{g=1} \textrm{log}(1-v_{gm}) \right).$$
\end{enumerate}

\section{Empirical Study}\label{simulations}

To illustrate the performance of the EIHD, we use data from the public use microdata files of the 2012 ACS, available for download from the United States Census Bureau (\url{http://www2.census.gov/acs2012_1yr/pums/}).  We construct a population of 842746 households from which  we sample $n=3000$ households comprising  $N=8686$ individuals. The sample includes households with two to six people, and  $\mathcal \{n_{2}^1, \dots, n_{6}^1\} = \{1541, 630, 525, 210, 94\}$. We work with the variables described in Table \ref{variable:description}, which mimic those in the U.\ S.\ decennial census. The structural zeros involve ages and relationships of individuals in the same household; see the supplementary material for a full list of the edit rules that we use. 

The Census Bureau purges the 2012 ACS public use microdata file of detectable errors. Therefore, for each household $i$, we treat its values on the public use file as error-free $\mathbf{X}_i^1$, and we purposefully introduce errors and missing values to the complete data file.  To do so, we randomly generate each $Z_i$, where $i=1, \dots, 3000$, from a Bernoulli$(\rho)$ distribution, where $\rho = 0.2$.  For each household with $Z_i=0$, we let $\mathbf{X}_i^1=\mathbf{Y}_i$.  For each household  with $Z_i = 1$, we sample error locations using (\ref{ErrorModel1}) and (\ref{ErrorModel2}), and sample reported values from (\ref{ReportingModel1}) and (\ref{ReportingModel2}). As we allow only detectable errors, we create errors only in variables used in the exemplary definitions of the structural zeros.  These include the gender and age of the household head, and gender, age and relationship to household head for the remaining household members.
 We set $\epsilon = (0.65, 0.80, 0.70, 0.85, 0.90)$ for these five variables.
For each  household  with $Z_i = 1$, we repeatedly sample values until the household fails the structural zero rules. 
 This results in approximately $17\%$ overall error rate for each variable across the 3000 sampled households. Although editing rates can be smaller in some contexts \citep{Jackson2010}, we view the $17\%$ error rate as a challenging but reasonable stress test for the EIHD.  Finally, we  introduce missing values for all variables, except household size, not subject to errors.  To do so, we randomly and independently blank $20\%$ of the values for each variable.

\begin{table}[t]
	\footnotesize
	\centering
	\caption{Description of variables used in empirical study. ``HH '' means household head.} \label{variable:description}
	\begin{tabular}[c]{p{2.4in}p{3.2in}}
		Description of variable & Categories \\ \hline 
		& \\
		\multicolumn{2}{l}{\underline{Household-level variables}} \\
		Ownership of dwelling & 1 = owned or being bought, 2 = rented\\
		Household size & 2 = 2 people, 3 = 3 people, 4 = 4 people, \\
		& 5 = 5 people, 6 = 6 people \\
		Gender of HH & 1 = male, 2 = female \\
		Race of HH & 1 = white, 2 = black,\\
		& 3 = American Indian or Alaska native, \\
		& 4 = Chinese, 5 = Japanese, \\
		& 6 = other Asian/Pacific islander, 7 = other race, \\
		& 8 = two major races, \\
		& 9 = three or more major races \\
		Hispanic origin of HH & 1 = not Hispanic, 2 = Mexican, \\
		& 3 = Puerto Rican, 4 = Cuban, 5 = other \\
		Age of HH & 1 = less than one year old, 2 = 1 year old,\\
		& 3 = 2 years old, \ldots, 96 = 95 years old \\
		& \\
		\multicolumn{2}{l}{\underline{Individual-level variables}} \\
		Gender & same as ``Gender of HH'' \\
		Race & same as ``Race of HH'' \\
		Hispanic origin & same as ``Hispanic origin of HH'' \\
		Age & same as ``Age of HH'' \\
		Relationship to HH & 1 = spouse, 2 = biological child, \\
		& 3 = adopted child, 4 = stepchild, 5 = sibling,\\
		& 6 = parent, 7 = grandchild, 8 = parent-in-law, \\
		& 9 = child-in-law, 10 = other relative, \\
		& 11 = boarder, roommate or partner, \\
		& 12 = other non-relative or foster child \\ \hline
	\end{tabular}
\end{table}


The method of generating the reported values implies that the true substitution probabilities $q_k$ in (\ref{ReportingModel1}) and (\ref{ReportingModel2}) do not equal $1/(d_k - 1)$ for all levels of variable $k$. For example, in the contaminated data that we generated, given that $E_{ijk} = 1$ for the relationship to household head variable, the probability of a spouse being wrongly reported as a parent is 0.177 whereas the probability of a spouse being wrongly reported as a sibling is 0.112. However, we still set $q_k = 1/(d_k - 1)$ when fitting our model to the data, mirroring a scenario where an agency uses a default application of EIHD and unknown true substitution rates.

We put the data for the household head as household level variables as suggested in \citet{AkandeEtAl2017b}. This offers computational gains relative to modeling the household head variables at the individual level. We refer readers to \citet{AkandeEtAl2017b} for details and simulation results showing that treating the household head variables as household level variables speeds up the sampler while generally maintaining the accuracy of estimands. For the cap-and-weight approach, we consider two choices for the weights, namely $(\psi_2,\psi_3,\psi_4,\psi_5,\psi_6) = (1,1,1,1,1)$, which corresponds to the original sampler, and  $(\psi_2,\psi_3,\psi_4,\psi_5,\psi_6) = (1/2,1/2,1/3,1/3,1/3)$, which is computationally efficient according to preliminary runs of the Gibbs sampler.

We run the MCMC sampler for 10,000 iterations, discarding the first 5,000 as burn-in and thinning the remaining samples every five iterations, resulting in 1,000 MCMC post burn-in iterates.  We set $F = 20$  and $S = 15$ based on initial tuning runs. For convergence, we examined trace and autocorrelation plots of $\alpha$, $\beta$, $\boldsymbol{\epsilon} = \{\epsilon_k: k=1,\ldots,p+q\}$ and a random sample of probabilities corresponding to univariate and bivariate distributions of the collected variables.  We derive these probabilities from (\ref{ModelSpecification1}) -- (\ref{ModelSpecification4}), averaging corresponding parameters over draws of the latent class indicators. 

The posterior number of occupied household-level clusters usually ranges from 10 to 14 for $(\psi_2,\psi_3,\psi_4,\psi_5,\psi_6) = (1,1,1,1,1)$ and from 13 to 19 for $(\psi_2,\psi_3,\psi_4,\psi_5,\psi_6) = (1/2,1/2,1/3,1/3,1/3)$, while the posterior number of occupied individual-level clusters across all household-level clusters ranges from 3 to 8 with $(\psi_2,\psi_3,\psi_4,\psi_5,\psi_6) = (1,1,1,1,1)$ and from 3 to 7 with $(\psi_2,\psi_3,\psi_4,\psi_5,\psi_6) = (1/2,1/2,1/3,1/3,1/3)$. Thus, it appears that $F=20$ and $S=15$ are adequate \citep{HuEtAl2018}. For each of the two choices of $\psi$, we create $L = 50$ multiply imputed datasets, $\textbf{Z} = (\textbf{Z}^{(1)}, \ldots, \textbf{Z}^{(50)})$, which are complete and error-free, from the posterior samples of $\mathcal{X}^1$. From the imputed datasets, we estimate the marginal probabilities of all the variables, bivariate probabilities of all possible  pairs of variables, and trivariate probabilities of all possible triplets of variables. There are 229 marginal probabilities, 18135 bivariate probabilities and 623173 trivariate probabilities.  We also estimate selected probabilities that depend on within-household relationships and characteristics of the household head. We combine the estimates using standard multiple imputation combining rules \citep{Rubin1987}. 

We only examine the performance of the multiple imputation inferences for the EIHD model.  We are not aware of any publicly available Fellegi-Holt editing systems for household level data.  For example, the ``editrules'' package \citep{DeJongeVanDerLoo2015} in $R$ applies Fellegi-Holt editing only for independent individuals. We also cannot easily compare to complete case analysis due to the rate of missingness.  About 80\% of the households have at least one missing entry, with the proportion rising to about 85\% when one adds the households with faulty data.

\begin{figure}[t]
	\centering
	\includegraphics[width=\textwidth, height = 3.1in, angle=0]{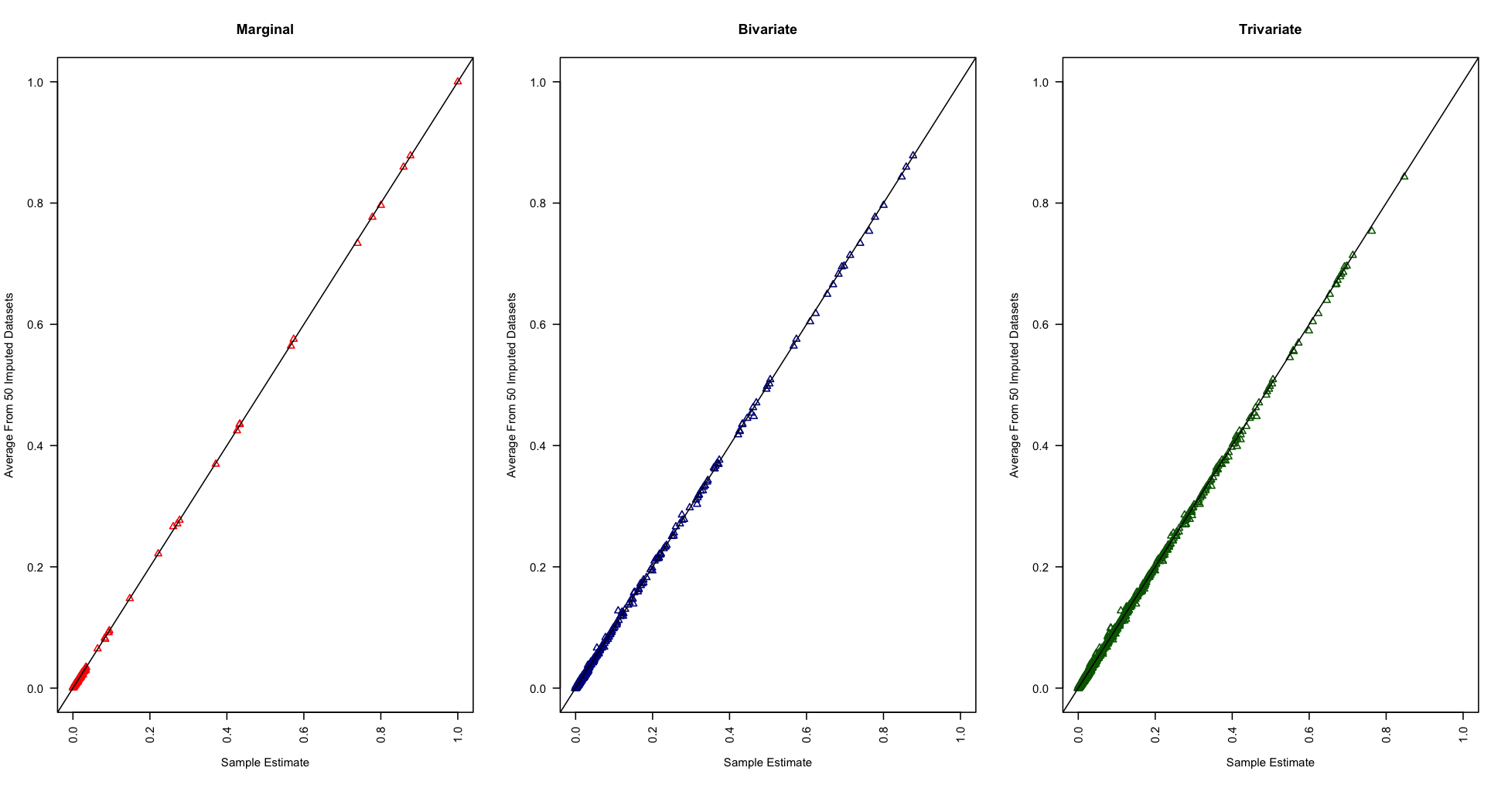}
	\caption{Marginal, bivariate and trivariate probabilities computed in the original and imputed datasets from the EIHD with $(\psi_2,\psi_3,\psi_4,\psi_5,\psi_6) = (1,1,1,1,1)$.}
	\label{AllProbs}
\end{figure}
\begin{figure}[t]
	\centering
	\includegraphics[width=\textwidth, height = 3.1in, angle=0]{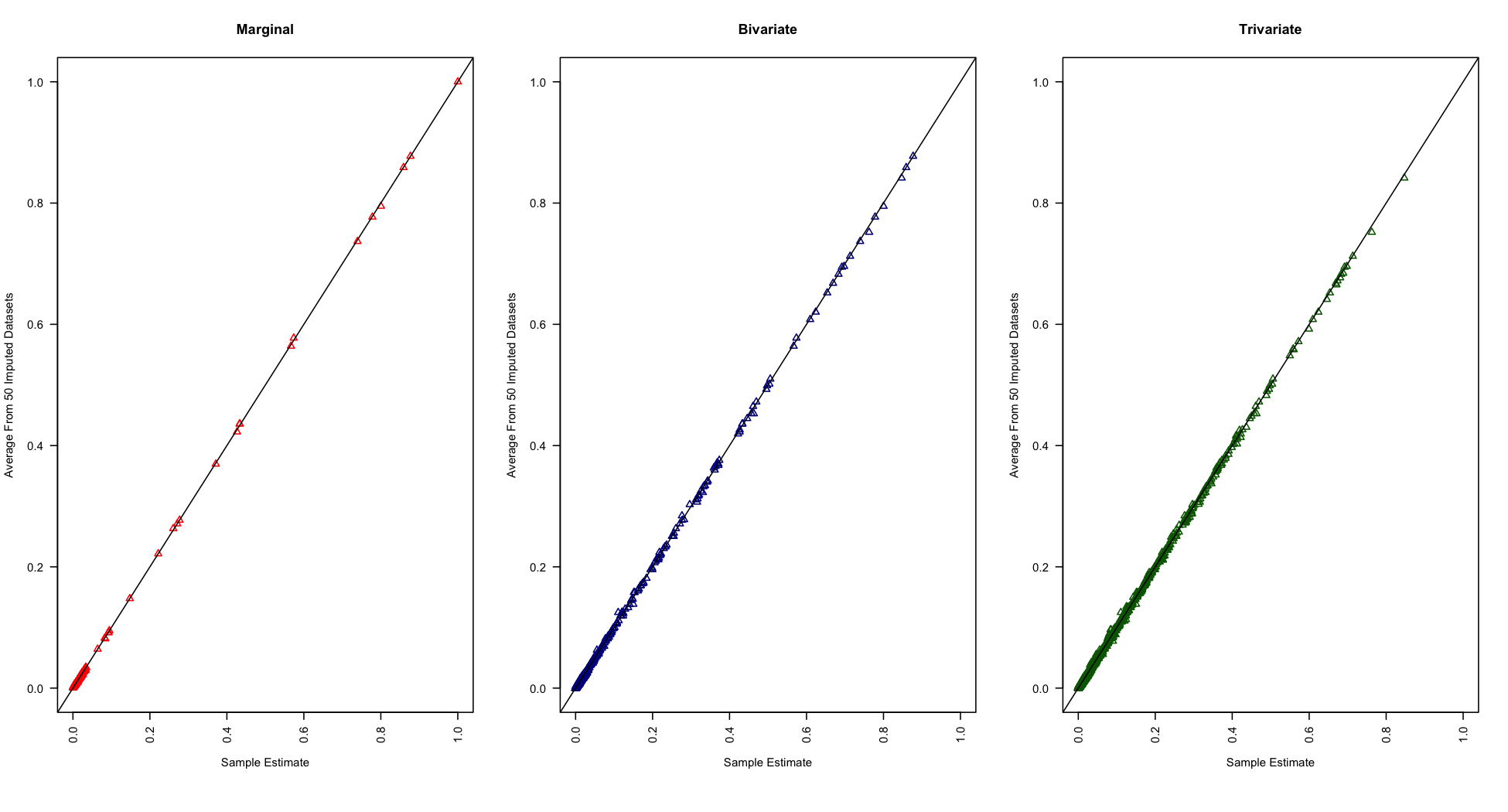}
	\caption{Marginal, bivariate and trivariate probabilities computed in the original and imputed datasets from the EIHD with $(\psi_2,\psi_3,\psi_4,\psi_5,\psi_6) = (1/2,1/2,1/3,1/3,1/3)$.}
	\label{AllProbs_Weighted}
\end{figure}

Figures \ref{AllProbs} and \ref{AllProbs_Weighted} display the estimated probabilities obtained from the multiple imputation combining rules 
for the EIHD with $(\psi_2,\psi_3,\psi_4,\psi_5,\psi_6) = (1,1,1,1,1)$ and $(\psi_2,\psi_3,\psi_4,\psi_5,\psi_6) = (1/2,1/2,1/3,1/3,1/3)$, respectively. In both versions of EIHD, the point estimates are very close to those from the original data (i.e., before the introduction of  missing and erroneous values), suggesting that the EIHD edit-imputations captured these features of the joint distribution of the variables. 
Table \ref{CI:n3000_gamma20} displays multiple imputation $95\%$ confidence intervals for the selected probabilities involving within-household relationships, as well as the values in the full population of 842746 households. 
For the most part, the intervals from both versions of EIHD are quite close to those based on the original data.  One exception is when estimating the proportion of households where 
everyone is the same race, especially for larger households. \citet{HuEtAl2018, AkandeEtAl2017b} also identified biases for the same estimands when using the NDPMPM to generate fully synthetic data or impute missing data.  They found that the bias gets smaller as the sample size increases. Estimating the proportion of households with couples whose age difference is less than 5 is also particularly challenging. This results because age can take 96 different values.  With so many levels, it is difficult to estimate within household relationshpis involving age with high accuracy. We revisit this issue in Section \ref{discussion}.

\begin{table}[t!]
	\footnotesize
	\centering
	\caption{Confidence intervals for selected probabilities that depend on within-household relationships in the original and imputed datasets. ``Sample'' is based on the sampled data before introducing errors and missing values, ``EIHD'' is the EIHD with $(\psi_2,\psi_3,\psi_4,\psi_5,\psi_6) = (1,1,1,1,1)$, and ``EIHD w/caps'' is the EIHD with $(\psi_2,\psi_3,\psi_4,\psi_5,\psi_6) = (1/2,1/2,1/3,1/3,1/3)$. ``HH '' means household head, ``SP'' means spouse, ``CH'' means child, and ``CP'' means couple. ``Population'' is the value in the full population of $842,746$ households.} \label{CI:n3000_gamma20}
	\begin{tabular}[c]{L{0.42\textwidth}|R{0.1\textwidth}R{0.12\textwidth}R{0.12\textwidth}R{0.12\textwidth}}
		& Population & Sample & EIHD & EIHD w/caps \\ 
		\hline
		All same race household: & & & & \\ 
		\hspace{8pt}  $n_i = 2$ & .942 & (.936, .958) & (.909, .938) & (.901, .934) \\ 
		\hspace{8pt}  $n_i = 3$ & .908 & (.863, .912) & (.830, .890) & (.820, .884) \\ 
		\hspace{8pt}  $n_i = 4$ & .901 & (.890, .938) & (.839, .906) & (.827, .894) \\ 
		\hspace{8pt}  $n_i = 5$ & .887 & (.848, .933) & (.772, .892) & (.770, .888) \\ 
		\hspace{8pt}  $n_i = 6$ & .871 & (.766, .914) & (.693, .875) & (.647, .846) \\ 
		SP present & .704 & (.687, .720) & (.682, .718) & (.682, .719)  \\ 
		Same race CP & .663 & (.645, .679) & (.628, .667) & (.624, .662) \\ 
		SP present, HH is White & .599 & (.590, .625) & (.585, .624) & (.585, .623)  \\ 
		White CP & .579 & (.572, .607) & (.563, .602) & (.561, .600) \\ 
		CP with age difference less than five & .494 & (.472, .508) & (.405, .442) & (.403, .441) \\ 
		At least one biological CH present & .490 & (.473, .509) & (.481, .519) & (.479, .517) \\ 
		Male HH, home owner & .472 & (.455, .491) & (.442, .482) & (.445, .485) \\ 
		HH over 35, no CH present & .416 & (.397, .432) & (.390, .428) & (.390, .428) \\ 
		HH older than SP, White HH & .320 & (.318, .352) & (.308, .345) & (.309, .347)  \\ 
		Adult female w/ at least one CH under 5 & .093 & (.078, .098) & (.077, .099) & (.077, .098) \\ 
		White HH with Hisp origin & .076 & (.059, .077) & (.055, .074) & (.055, .074)  \\ 
		Non-White CP, home owner & .063 & (.048, .064) & (.039, .056) & (.039, .057)  \\ 
		Two generations present, Black HH & .059 & (.043, .059) & (.045, .063) & (.045, .063) \\ 
		Black HH, home owner & .052 & (.039, .053) & (.038, .054) & (.038, .054)  \\
		At least three generations present & .041 & (.032, .046) & (.032, .047) & (.031, .047)\\ 
		SP present, HH is Black & .041 & (.028, .042) & (.028, .043) & (.028, .043)  \\ 
		At least two generations present, Hisp CP & .040 & (.028, .040) & (.027, .041) & (.027, .041) \\ 
		Hisp CP with at least one biological CH & .038 & (.027, .040) & (.026, .040) & (.026, .049)  \\ 
		White-nonwhite CP & .035 & (.027, .039) & (.032, .050) & (.034, .052) \\ 
		One grandchild present & .032 & (.021, .032) & (.023, .038) & (.024, .038)  \\ 
		Hisp HH over 50, home owner & .030 & (.023, .035) & (.023, .036) & (.023, .038)\\ 
		Adult Black female w/ at least one CH under 18 & .030 & (.022, .034) & (.020, .033) & (.020, .033) \\
		Adult Hisp male w/ at least one CH under 10 & .027 & (.015, .025) & (.014, .024) & (.014, .025)  \\
		At least one stepchild & .026 & (.021, .032) & (.020, .033) & (.020, .033)  \\  
		Only one parent & .023 & (.015, .025) & (.018, .030) & (.017, .030)  \\  
		Three generations present, White CP & .013 & (.008, .016) & (.007, .016) & (.007, .016)  \\ 
		At least one adopted CH, White CP & .010 & (.008, .015) & (.007, .015) & (.007, .016) \\ 
		Black CP with at least two biological children & .009 & (.004, .010) & (.005, .012) & (.005, .012)  \\ 
		Black HH under 40, home owner & .006 & (.003, .008) & (.005, .014) & (.005, .013) \\ 
		White HH under 25, home owner & .003 & (.000, .002) & (.002, .009) & (.001, .007) \\ 
		\hline
	\end{tabular}
\end{table}

The supplement includes results from additional simulations based on other measurement error models and missing data rates. In particular, we examine simulations with (i) the same substitution model but a higher rate $\rho=0.4$ of households that violate the constraints, and $30\%$ missing data, (ii) a non-uniform substitution model with fixed error rates where $\rho=0.2$, and $20\%$ missing data, and (iii) a non-uniform substitution model with Beta distributed error rates where $\rho=0.2$, and $20\%$ missing data. Overall, the performance of the EIHD is similar qualitatively to what we present here.   

\section{Discussion} \label{discussion}

Simultaneously estimating multivariate relationships accurately, capturing within-household relationships, adjusting for measurement errors, and respecting structural zeros in estimation and imputation is a challenging task.   The simulation results here suggest that the EIHD does a  good job at that task, at least when the measurement error modeling assumptions are approximately true.  As with any imputation strategy applied on genuine data, it does not capture all associations perfectly.  In particular, quantities that involve many individuals within the same household, such as combinations of races, or that are based on many categories, such as multivariate probabilities involving ages, can be difficult to estimate.  In large part this is a result of inadequate sample size for the model to capture such quantities in the larger households. For example, our sample data contains 1541 households of size two but only 94 households of size six.  One possible solution is to redefine variables with many categories to reduce the number of parameters. For example, if acceptable one can use age intervals rather than discrete values of age, or possibly replace age with a variable capturing the difference from the household head's age. 

The EIHD can be computationally expensive due to the rejection sampling steps. It can take time to generate feasible imputations for the faulty households, especially when the true error rates are high or when the proportion of households with detectable errors is high. Fortunately, the rejection sampling step can be easily parallelized. 
 This should  speed up the sampler by a factor of roughly the number of processors available. The rejection steps are very expensive for large households sizes (e.g., 10 or more people) because the size of $\mathcal{S}_h$ grows exponentially in $h$. For such households, which usually are not present in large numbers, we suggest exploring ad-hoc versions of the EIHD; for example,
imputations for each large household with faulty values can be generated from a fixed set of proposals generated using a variation of hot-deck or cold-deck imputation.

The EIHD model can be used to provide disclosure limitation in public release files.  In particular, it can be an engine for creating synthetic data \citep{RaghunathanEtAl2003, Reiter2005}, enabling agencies to handle the disclosure protection and edit-imputation in one integrated approach \citep{KimEtAl2015a, KimEtAl2018}. Simply, once the agency has draws of the EIHD model parameters estimated with the faulty data, the agency uses the rejection sampler to generate the synthetic households following the steps in \citet{HuEtAl2018}.  

The EIHD model also can be used in conjunction with the disclosure control technique PRAM, also known as the post randomization method \citep{GouweleeuwEtAl1998}.  In PRAM,  the agency purposefully introduces measurement errors to categorical values using what is essentially the measurement error model from  Section \ref{measurementerror} with fixed $\epsilon_k$.  To illustrate an application of PRAM, for each individual $(ij)$, we keep each $Y_{ijk}$ at its collected value with probability $0.6$ and reset $Y_{ijk}$ to a random draw from the other $d_k-1$ values in variable $k$ with probability $0.4$.  Agencies first could use PRAM to perturb confidential values, then use the EIHD synthetic data engine with $\epsilon_k$ fixed at the PRAM probabilities (0.6 in this example) and estimated with the perturbed data.  The resulting synthetic datasets would satisfy all edit constraints as well as propagate uncertainty due to the perturbations and synthesis.  This integration of PRAM and EIHD also generates synthetic household data that satisfy differential privacy \citep{Dwork2006}, since PRAM applied to all the variables satisfies differential privacy---albeit with a privacy budget that scales with the dimension of the table---and applying the EIHD synthesis engine to the perturbed data is a post-processing step that does not negatively affect the privacy guarantees of PRAM.  We leave investigation of the use of EIHD for disclosure limitation as a topic for future research.

Finally, as with all empirical evaluations of new methods, the results presented here are based on limited simulations. In particular, as seen in related work in \citet{KimEtAl2015b}, we expect nonignorable missingness or error mechanisms and more generally, severe model misspecification, to degrade the performance of the EIHD compared to the presentation here.  More informative measurement error models are necessary for any method, including EIHD and Fellegi-Holt approaches, to be effective for such mechanisms.  An important future research topic is to incorporate nonignorable measurement errors in the EIHD approach, as well as to assess the sensitivity of inferences from the completed datasets to different specifications of the measurement error models.

\section{Supplementary Materials}
The supplementary material contains a list of the structural zero rules used in fitting the EIHD model and the additional empirical studies referred to in Section \ref{simulations}.

\renewcommand{\refname}{References}
\printbibliography
\end{document}